# Stereo-specific internally entangled roaming mechanism in the reaction of unstable $B_5$ cluster with H

Takehiro Yonehara[1*], Takahito Nakajima[1], Hikaru Kobayashi[2], King-Chuen Lin[3,4], Toshio Kasai[2,3]

[1]RIKEN Center for Computational Science, Kobe 650-0047, Japan

[2]Institute of Scientific and Industrial Research, Osaka University, Ibaraki, Osaka 567-0047, Japan

[3]Department of Chemistry, National Taiwan University, Taipei 10617, Taiwan

[4]Institute of Atomic and Molecular Sciences, Academia Sinica, Taipei 10617, Taiwan

*To whom correspondence should be addressed should be addressed at tkyn2011@gmail.com or takehiro.yonehara@riken.jp

**Abstract**

A new type of, called "stereo-specific entangled", roaming mechanism is presented in the reaction of structurally unstable $B_5$ cluster with the hydrogen atom, by using the direct *ab initio* trajectory calculation with a practical level of DFT method using the rang-separated functional. In this mechanism, the light moiety, i.e. H roams around far from the $B_5$ cluster caused by a stereo-specific energy transfer of translational energy to the cluster vibrational energy through the direct coupling due to the large amplitude motion of the unstable $B_5$ cluster. We confirmed by the comparative computation that a clear change occur from the roaming trajectory to the repulsive trajectory, when we froze the vibrational motion of the $B_5$ cluster. This drastic change of the trajectory due to the change between the structurally unstable $B_5$ cluster and the frozen $B_5$ cluster revealed the importance of kinematic energy transfer resulted in the soft-landing-like hydrogen absorption which manifested the present stereo-specific entangled roaming mechanism. The trajectory calculation revealed that the present new type of roaming mechanism is very sensitive to the azimuthal angle of the H atom attack, especially to the bisection of the B-B bond of the $B_5$ cluster, where the energy flow from relative kinetic energy to the vibrational energy of the cluster efficiently takes place through the roaming trajectory driven due to the structural fluctuation of the $B_5$ cluster.

Table of contents

I.  Introduction

The large amplitude motion of clusters plays an important role in chemical and physical processes in many complex systems, such as self-organization of surface structure in chemical vapor depositions, multi-channel reactive and inelastic dynamics in the gas phase, also reaction dynamics at surfaces in many catalytic processes at the nano-scale. The knowledge of these systems provides us relevant know-how to develop energy-conversion devices in industrial aspect like solar cells.

The ground state of boron clusters is characterized by the existence of several local minima associated with a variety of bonding pattern due to the electron deficiency in a 2p valence orbital space of the B atom, i.e. $(1s)^2(2s)^2(2p)$.[1] The deformation of boron cluster offers us additional reactivity, but its mechanism is not fully understood because their conformational changes in chemical dynamics are complicated.[1] Boron clusters and crystals possess a structural stability at the ground state minimum in addition to many local minima with low energy barriers on the potential energy surface and especially bond frustration occurs due to electron deficiency in the valence space. This is intuitively attributed to an explosive number of combination patterns in the electron filling of vacant valence orbitals associated with the aggregation of boron atoms.[2] Because of such rich vacancy in a 2p valence space, but not in the d-orbitals, we refer the boron element to semi-metal at a low position in the periodic table and it offers us a curious and interesting chemistry in material world. Many kinds of bond alternations and surgeries are mediated by this atom.[3] We recently have investigated boron clusters from a viewpoint of radical fluctuation in non-adiabatic electron dynamics with the manifold of quasi-degenerate excited electronic states coupled with vibronic motion of the cluster system, and proposed a new type of complex reaction dynamics through many non-adiabatic transition paths.[4,5] Li analyzed symmetry breaking dynamics in nonadiabatic electron dynamics in $B_4$ and proposed the selective transition dynamics associated with nonadiabatic coupling through the vibrational mode involved with the conical intersection.[6] Arasaki and Takatsuka have made a more precise and comprehensive analysis on the bond fluctuation in non-adiabatic electron dynamics.[7]

The concept of the transition state (TS) has been considered as a solid foundation in bi-molecular uni-molecular reactions and photodissociations. We performed works on the isomerization and quantum effect on multi-channel dissociation reactions of formaldehyde molecule by using a classical as well as quantum dynamics calculation under the energy region without roaming process.[8-10] In 2004, the existence of roaming process was first confirmed in the same photodissociation of formaldehyde $H_2CO \rightarrow H_2 + CO$ by Suits and Bowman groups.[11] The roaming trajectories are characterized by a large amplitude of motion of a small moiety in a region which is far from the other part of the system without dissociation. The conventional transition state theory cannot describe such roaming dynamics.[11-21] Nowadays, the roaming dynamics in chemical reactions is recognized to be ubiquitous in many reactions.[22] They include (1) dissociation pathways with loose saddle points (SP), found in aldehydes, acetone, and formic[23-25,18], (2) conical intersections involving multiple state[26,27], (3) isomerization-mediated paths, found in $NO$[28], and (4) bimolecular analogues showing the roaming feature[29], also found in reactions such as $OH+HBr \rightarrow H_2O+Br$.[30] Accordingly, the concept of roaming mechanisms becomes more complicated and diverse than ever expected a decade ago. It is therefore important to investigate theoretical origins of the roaming behaviors in order to classify and simplify the types of roaming mechanisms.

For this reason, we employed in the present study a small size of unstable boron cluster $B_5$ with the $C_{5v}$ symmetry as the initial cluster structure in order to enhance any effect of kinematic coupling between $B_5$ and the incident hydrogen atom. We were able to clarify that the incomplete energy transfer from collision energy into internal energy of the cluster enhances absorption of the hydrogen atom to the cluster enhances followed by repulsive dynamics accompanied with a roaming trajectory that creates a hydrogen migration dynamics with a large amplitude of motion. We confirmed this new type of roaming mechanism in the $B_5$ + H reaction by using the direct *ab initio* trajectory calculation with a practical level of DFT method with the range separated functional, CAM-B3LYP/6-31G(d).[31]

In the following, we provide the details on the initial conditions of the reaction system of $B_5$ + H and the explanation of the methods for the trajectory calculation in section II. The results are discussed in Section III, followed by the concluding remarks in the final section IV.

## II. Reaction system and the initial conditions for trajectory calculation

=================
**Figure 1**
=================

Figure 1 shows the canonical Kohn-Sham molecular orbitals from HOMO-2 to LUMO for the initial referential geometry of the $B_5$-H system. These orbitals were obtained by a CAM-B3LYP/6-31G(d) level calculation, where $B_5$ has unstable reference geometry with $C_{5v}$ symmetry and the H atom is located far from the $B_5$ cluster. We used a moderate distance between two moieties of $B_5$ and H of $B_5$H due to a convergence problem especially in this one-point energy calculation. Since this system consists of 26 electrons as the total, thus the 11-, 12-, 13- and 14-th orbitals are plotted in the canonical order as (a), (b), (c) and (d) in the figure. We included the H atom in the MO calculations because the reaction system conserves the consistent presentation of the $B_5$ + H system in the *ab initio* dynamics calculation. As we see in panel (c), the HOMO orbital has dominant amplitude toward the H direction and a sufficient negative value of LUMO energy gives rise to an inherent feature of the hydrogen absorption in the potential energy. However, this initial geometry is not sufficient to predict reactivity in the $B_5$ + H reaction, because the lack of stereo-specificity factor affects the reactivity especially in the roaming type of hydrogen absorptive dynamics which will be discussed in detail later. We employed the second order symplectic integrator[32,33] for the time propagation with a 0.1 *fs* time increment. To initiate chemical reaction, some relative collisional velocity was given at the initial time in the simulation, while the kinetic energy of the initial internal motion of $B_5$ boron cluster was set to be zero. Thus the resultant total momentum and angular momentum become zero, for we assume the line-of-center collision in this trajectory simulation. The symbol of $E_{rel}$ is used as kinetic energy evaluated from this relative collisional velocity.

=================
**Figure 2**
=================

Figure 2 represents the spatial coordinates which define the positons of the six atoms in the $B_5$ + H reaction. The hydrogen atom H is shown with the red circle and the five boron atoms are shown with the blue circles. The $B_5$ cluster locates on the XY plane holding the mirror symmetry with respect to the Y axis and the length of each B-B bond is 1.56 Å. The azimuthal and polar angles are given as α and β, respectively, which define the approaching direction of the H atom toward the $B_5$ cluster. We assume in the present computation that a collision takes place along the line of center of the H atom and the center of mass of the $B_5$ cluster, therefore the total angular momentum is kept as zero during collision. For simulating classical trajectories, we employed the CAM-B3LYP/6-31G(d) level of the *ab initio* electronic structure calculations with the spin restricted the spatial orbitals and the neutral charge is assumed for the whole system. We evaluated the potential gradient as the function of time, by use of GAMESS program suite[34] with the modification about time integrator.

The conformation of boron cluster was initially set as regular pentagonal with the B-B length of 1.56 Å. This $B_5$ structure is known to be neither optimal nor stationary on its ground state potential energy surface. Zhai *et al.* simulated by using B3LYP/6-311+G* level calculation that the optimized $B_5$ geometry to be planar with $C_{2v}$ symmetry with three three-center bondings in the ground electronic state.[35] However, we employed pentagonal symmetry as the initial geometry of $B_5$ in the present study because pentagonal geometry offers us more efficient and clear presentation for the kinematic coupling between the mutual collisional motion of H and the $B_5$ cluster internal dynamics, to which we especially should like to pay attention in this work.

## III. Results and discussion

The essential behaviors of trajectories up to the first 100 *fs* are presented in sections A and B, for introducing the stereo-specific entangled roaming mechanism, which we computationally observed in the present work, especially it occurs strongly associated with the coupling of the H atom translational energy to the vibrational energy of the structurally unstable $B_5$ cluster.

### A. The behavior of the H atom and the $B_5$ cluster trajectories upon the polar angle β when the H atom directly approaches to one B atom of $B_5$ at the azimuthal angle of α = 54°

The radial trapping of two interacting groups with a large angular momentum would be likely occur due to dynamical centrifugal barrier as the result of the energy conservation for the whole system. The stereo-specific entangled roaming dynamics in this work would be an analogous manifestation of this radial trapping via kinematic energy transfer to the vibrational energy of the unstable $B_5$ cluster. Note that we do not include total rotational motion in this collision dynamics calculation for we assume a collision takes place along the line of centers in the reacting system. In order to clarify such strong kinematic coupling during collision, we chose the unstable pentagonal structure of the $B_5$ cluster as the initial structure of the $B_5$ cluster in the present calculation, but the B-B bond of the cluster is easily deform when the H atom approaches to the $B_5$ cluster.

==================
**Figure 3**
==================

Figure 3 shows typical four trajectories as the variation of the polar angle β = 5°, 15°, 35°, 65° in the case when the H atom approaches to $B_5$ at the azimuthal angle α = 54°. This azimuthal angle corresponds to the angle of H attack toward the bisection of the B-B bond of the cluster, which corresponds to the belly (maximum amplitude of vibrational motion of the bond). The number attached to each small circle of the trajectory stands for the initial spatial position of H with the polar angle, namely β = 5°, 15°, 35°, 65°, respectively, in the line-of-center collision between H and $B_5$. The four curves connecting to the four initial positions shown in red, green, yellow, and black denote the trajectories of the H atom. The rather congested trajectories of the five B atoms are plotted in blue. The incident relative collision energy was set to $E_{rel}$ = 1.088 eV, so that the energy of the H atom trajectory would asymptotically reach to the dissociation energy. Panel (A) shows the trajectories in full dimensional space and its projection onto XY, XZ, or YZ plane is shown in panel (B), (C), or (D), respectively. The shape of each H trajectory varies as the polar angle $\beta$ changes, but it shows a similar behavior like as a long-lived collision complex of H-$B_5$ or a roaming type.

**B. The behavior of the H atom and the $B_5$ cluster trajectories upon the polar angle $\beta$ when the H atom approaches to the bisection of the B-B bond of $B_5$ at the azimuthal angle of $\alpha = 18°$**

==================
**Figure 4**
==================

Similarly, Figure 4 shows typical four trajectories as the variation of the polar angle $\beta = 5°, 15°, 35°, 65°$ in the case when the H atom approaches to $B_5$ at the azimuthal angle $\alpha = 18°$. This azimuthal angle corresponds to the angle of H attack directly toward one of the B atoms of the cluster. The number attached to each small circle of the trajectory stands for the initial spatial position of H with the polar angle $\beta = 5°, 15°, 35°, 65°$, respectively, same as in Figure 3 and the four initial positions of the H trajectory is shown in red, green, yellow, and black. The congested trajectories of the five B atoms are plotted in blue. The incident relative collision energy was the same as $E_{rel} = 1.088$ eV. Panel (E) shows the trajectories in full dimensional space and its projection onto XY, XZ, or YZ plane is shown in panel (F), (G), or (H), respectively. The shape of the H trajectories as the function of the polar angle $\beta$ changes looks similar to be the direct rebound scattering.

The comparison of two different types of trajectory behaviors at two different azimuthal angles $\alpha = 54°$ in Figure 3 and $\alpha = 18°$ in Figure 4 suggests a clear stereo-specificity in the $B_5$ + H reaction. When $\alpha = 54°$, the H atom is incident to the bisection of the B-B bond of the cluster then the H atom roams around the $B_5$ cluster, while the B atoms of the cluster tremble to some extent in a synchronized way with the H atom roaming as seen in panel (A). On the other hand, when $\alpha = 18°$, the H atom directly attacks the B atom of $B_5$ and the resulting trajectories of the H atom show the clear direct rebound scattering behavior directly experiencing repulsive force as seen in panel (E). As for the polar angle $\alpha$ variation for the initial position of the H atom, it seems to be no such drastic change in contrast to the case of the azimuthal angle $\beta$ variation.

C. **The effect of structural instability of the $B_5$ cluster upon the reactivity of the $B_5$ + H reaction**

================
**Figures 5 and 6**
================

In order to make it clear any effect of the structural instability of the $B_5$ cluster upon the reactivity of the $B_5$ + H reaction or the trajectory behavior, we performed the competitive trajectory calculations under the same initial conditions given in Figures 3 and 4 except for the following condition. We froze the vibrational motion of the $B_5$ cluster in Figures 5 and 6, namely we set the all B atoms at the fixed spatial positions all the time. As seen from the trajectories in panel (I) of Figure 5, set at the azimuthal angle $\alpha$ = 54°, the roaming type trajectories of panel (A) completely changed to the repulsive trajectory pattern. The repulsive type trajectories in panel (M) at the azimuthal angle $\alpha$ = 18° in Figure 6, however, a similar repulsive pattern is almost reproduced even if we froze the vibrational motion of the $B_5$ cluster. This result in Figure 5 and Figure 6 strongly suggests that there would be no coupling of the H atom translational motion with the internal motion of the B5 cluster when the H atom attack on top of the B atom the B-B bond. In contrast, only the H atom attacks toward the bisection of the B-B bond of the cluster strongly coupled with the vibrational motion of $B_5$ and leads to the present roaming mechanism. Thus, it is a clear signature that the new type of roaming mechanism, which we observe in this study should be induced in a kind of synchronized way through coupling between the radial motion of the H atom and the internal motion of $B_5$, or clear kinetic energy transfer of the H atom to the structurally unstable $B_5$ cluster. It would be a reason because the H atom attack at the belly of the B-B bond of the cluster should couple more efficiently with the cluster vibrational motion than the direct H atom attack at the B atom of the cluster. For confirming this new type of roaming mechanism, we performed the long-time period simulation using the long-time interval trajectory calculation up to 500 *fs* in the next section.

D. **The long-time interval trajectory calculation and the time dependence of the kinetic energies of the H + $B_5$ system.**

================
**Figures 7**
================

================

The stereo-specific entangled roaming trajectories were observed by the grid search in the previous trajectory calculations up to 100 *fs*. In this section, it is worthwhile to look at the overall behavior of the roaming trajectory for longer time interval. We thus performed the long-time interval calculation up to 500 *fs* to ensure any different characteristics of the present roaming mechanism to the ordinary roaming mechanism. The initial conditions were set as ( $R$, $\alpha$, $\beta$, $E_{rel}$ ) = ( 5.5, 72°, 5°, 1.088 ), where $R$ is the distance between the H atom and the $B_5$ cluster in the unit of Å at $t = 0$ and $E_{rel}$ is the initial collision energy in the unit of eV. $\alpha$ and $\beta$ are the azimuthal and the polar angles, respectively. The H atom approach along the azimuthal angle $\alpha = 72°$ in Figure 7 which is a little larger angle than $\alpha = 54°$ in Figure 3, but it is basically a similar angle to the bisection of the B-B bond of the cluster. The polar angle $\beta = 5°$ corresponds to almost the Z direction. Panel (a) shows the trajectory for the full dimensional collision dynamics and panel (b) is its projection onto the YZ plane.

Panel (c) in the figure displays the time-dependence of the following two quantities with the red solid line and the blue dashed line, are related to the kinetic energies defined in the following equations (1) and (2).

Red solid line:

(1) $$E^{Kinetic}_{relative} \equiv \frac{1}{2}\mu \vec{V}^2$$

where $E^{Kinetic}_{relative}$ is the relative kinetic energy of the H + $B_5$ system, and V is the difference velocity between the velocity of $B_5$, $V^{COM}_{[B5]}$ and that of the H atom $V^{COM}_H$, in which COM stands for the center of mass of system. Because of the mass difference, $V^{COM}_H$ is much faster than $V^{COM}_{[B5]}$.

(2) $$\mu \equiv \frac{M_H M_{[B_5]}}{M_H + M_{[B_5]}} \quad \vec{V} \equiv \vec{V}^{COM}_{[B_5]} - \vec{V}^{COM}_H$$
$$\vec{V}^{COM}_{[B_5]} \equiv \frac{1}{5}\sum_{i=1}^{5}\vec{v}_{[B(i)]} \quad \vec{V}^{COM}_H \equiv \vec{v}_H$$

Blue dashed line:
On the other hand, $E^{Kinetic}_{relative[B5]}$ is defined as the sum of the kinetic energy of the five boron atoms as given in eq.(3).

(3) $$E^{Kinetic}_{internal:[B_5]} \equiv \sum_{i=1}^{5}\left\{\frac{1}{2}M_B(\vec{w}_{B(i)})^2\right\} \quad \vec{w}_{B(i)} \equiv \vec{v}_{B(i)} - \vec{V}^{COM}_{[B_5]}$$

Where $B(i)$ is the i-th boron atom in the $B_5$ cluster, and $w_{B(i)}$ is the velocity difference between the velocity of i-th B atom and the velocity of the cluster in the center of mass.

We label the key time points along the trajectory of the H atom in panel (a), in order to compare the time dependences of the relative energy (in red), which is mostly carried by the H atom, and the internal kinetic motions of the sum of five B atoms of the cluster (in blue) as seen in panel (c). The first impact of the hydrogen atom to the cluster at ~20 *fs* (the sharp dip in red) causes a significant energy transfer with ~1 eV from the relative collision energy to the internal energy of the $B_5$ cluster. Soon after this impact, the H atom exerts a strong repulsion followed by the lingering roam motion of the H atom at far distance about 6 Å from the center of $B_5$ from t = 50 to 280 *fs* with small relative kinetic energy. This trajectory of the H atom with small kinetic energy corresponds to a regular type of roaming motion trapped by a long attractive force between H and $B_5$. The secondary approach of the H atom to the $B_5$ cluster is seen around t = 280 *fs*. As seen clearly in panel (c) as well as panel (a), this secondary soft impact of the H atom features quite differently from the first impact. The entangled strong oscillations occur as displayed in red and blue curves for the two moieties H and $B_5$ in a synchronized way. This cooperating strong oscillation of the two moieties is a manifestation of the rapid exchange back and forth between kinetic energy and potential energy in this case vibration and internal motion of the collision complex during the period from 300 to 450 *fs*. Therefore, according to the observation of roaming motion following the first and second impact involved with complicate energy exchange between the cluster and hydrogen atom, we may call such oscillation having large amplitude of motion as the entangled roaming motion.

In contrast, when we freeze the all B atoms of the cluster during collision with the H atom, we only observe a direct rebound scattering in the full dimensional dynamics in panel (d) as well as in panel (e), i.e. its projection onto the YZ plane. In the time dependence of the kinetic energy in panel (c), we observe the repulsive impact feature again between H and $B_5$ at ~20 *fs*. However, there is no oscillating feature of the whole system can be seen. Note that the blue dotted straight line simply shows that all B-B bonds are frozen during the collision with the H atom. The red curve, on the other hand, represents that the H atom flies away from the potential well produced by the interaction between H and $B_5$. This result strongly rationalizes the fact that the roaming mechanism found in this study can be classified as a new type one we may be called as entangled roaming mechanism.

Thus far, we have obtained the numerical evidence that the stereo-specific entangled roaming mechanism specifically occurs in the collision dynamics of the structurally instable $B_5$ cluster and the hydrogen atom. Analogous phenomenon is expected to find in systems related to the phase-transition dynamics of complex chemical reactions, where roaming type of interactions might play a key role in such many-body entangled systems. Further experimental confirmations of the present new type of roaming dynamics would shed light on the variety of new features in chemical reactions in general.

**IV. Concluding remarks**

The present report presented by use of the *ab initio* direct trajectory calculation a new type of roaming mechanism caused by the stereo-specific entangled energy transfer in the reaction of structurally unstable $B_5$ cluster with the hydrogen atom. In this mechanism, an efficient energy transfer from translation to the cluster vibrational motion occurs via the kinematic coupling under the stereo-specific collisional conditions. When the hydrogen atom approaches toward the bisection of the B-B bond of the $B_5$ cluster, the trajectory causes the large amplitude of motion of the whole $B_5$ + H system. The comparison between the trajectories with the structurally unstable $B_5$ cluster and that with the frozen $B_5$ cluster up to 500 *fs* revealed the importance of kinematic energy transfer due to the soft-landing-like hydrogen absorption which manifested the present stereo-specific entangled roaming mechanism. The present computational findings of the new type of roaming mechanism would provide a potential hint to interpret interesting phenomena occurring in surface reaction dynamics as well as in condensed phase chemical dynamics.

**Acknowledgements**

T.Y. thanks MEXT because this research was supported by MEXT, Japan, ``Next-Generation Supercomputer Project'' (the K computer project) and ``Priority Issue on Post-K Computer'' (Development of new fundamental technologies for high-efficiency energy creation, conversion/storage and use). Some of the computations in the present study were performed using the Research Center for Computational Science, Okazaki, Japan, and also HOKUSAI system in RIKEN, Wako, Japan. Authors deeply appreciate Prof. Takatsuka teaching us the importance and uniqueness of boron cluster in chemistry. T.K. thanks financial supports from the Ministry of Science and Technology (MOST), Taiwan and also from National Taiwan University.

**Corresponding Author**

*Email for T. Yonehara: tkyn2011@gmail.com   ( or takehiro.yonehara@riken.jp )

**ORCID**

T. Yonehara: https://orcid.org/0000-0003-4189-9414

**Ethics Declaration**

The authors declare no conflict of interest.

**Figure Captions**

**Figure 1** Canonical molecular orbitals. In the conformation of the $B_5$ cluster is set as $C_{5v}$ symmetry. The H atom is placed on the vertical symmetry plane including the $B_{(1)}$ atom. The distance of center of masses of two moieties $B_5$ and H is 5.7 Å. Panels (a), (b), (c) and (d) display isosurfaces of quantum amplitude for HOMO-2, HOMO-1, HOMO and LUMO of the whole system, respectively. MO energies of (a-d) correspond to -9.987, -9.497, -2.857 and -2.558 eV, respectively. Calculation level is the same as that used in dynamics calculation.

**Figure 2** Schematics of the initial relative geometry of the $B_5$ cluster and the H atom. The red circle represents the H atom atom and the blue circles correspond to five B atoms of the cluster. The $B_5$ cluster with regular pentagonal shape locates initially on the XY plane with the symmetry axis of Y and the length of each edge is 1.56 Å. Here, we use $\alpha$ and $\beta$ as the azimuthal and the polar angle, respectively for defining the initial direction of the H atom approach to $B_5$ in the Cartesian coordinate.

**Figure 3** Four representative trajectories for the first 100 *fs* at the azimuthal angle α is 54°, which feature a roaming-type mechanism in the $B_5$ + H reaction. The trajectories of the H atom presented with the numbers attached at the initial positions with the small circles, i.e. 5, 15, 35, and 65, respectively. These numbers correspond to the polar angle β for the initial positions. Five trajectories in blue represent the B atom of the $B_5$ cluster movements for 100 *fs*. Incident relative collision energy was set to $E_{rel}$ = 1.088 eV. Panel (A) shows the full dimensional dynamics. Panels (B), (C), and (D) are these projections onto the XY, XZ and YZ planes, respectively. The units of three coordinate axes are given in Å.

**Figure 4** Four representative trajectories for the first 100 *fs* at the azimuthal angle α is 18°, which all feature a rebound scattering mechanism in the $B_5$ + H reaction. The trajectories of the H atom presented with the numbers attached at the initial positions with the small circles, i.e. 5, 15, 35, and 65, respectively. These numbers correspond to the polar angle β for the initial positions. Five trajectories in blue represent the B atom of the $B_5$ cluster movements for 100 *fs*. Incident relative

collision energy was set to $E_{rel}$ = 1.088 eV. Panel (E) shows the full dimensional dynamics. Panels (F), (G), and (H) are these projections onto the XY, XZ and YZ planes, respectively. The units of three coordinate axes are given in Å.

**Figure 5**    Four representative trajectories for the first 100 *fs* at the azimuthal angle α is 54°, in which we froze the vibrational motion of the $B_5$ cluster and set the all B atoms at the fixed spatial positions. The trajectories of the H atom presented with the numbers attached at the initial positions with the small circles, i.e. 5, 15, 35, and 65, respectively. These numbers correspond to the polar angle β for the initial positions. Five trajectories in blue represent the B atom of the $B_5$ cluster movements for 100 *fs*. Incident relative collision energy was set to $E_{rel}$ = 1.088 eV. Panel (I) shows the full dimensional dynamics. Panels (J), (K), and (L) are these projections onto the XY, XZ and YZ planes, respectively. The units of three coordinate axes are given in Å. The roaming type trajectories in Figure 3 changed to complete repulsive trajectories.

**Figure 6**    Four representative trajectories for the first 100 *fs* at the azimuthal angle α is 18°, in which we froze the vibrational motion of the $B_5$ cluster and set the all B atoms at the fixed spatial positions. The trajectories of the H atom presented with the numbers attached at the initial positions with the small circles, i.e. 5, 15, 35, and 65, respectively. These numbers correspond to the polar angle β for the initial positions. Five trajectories in blue represent the B atom of the $B_5$ cluster movements for 100 *fs*. Incident relative collision energy was set to $E_{rel}$ = 1.088 eV. Panel (H) shows the full dimensional dynamics. Panels (N), (O), and (P) are these projections onto the XY, XZ and YZ planes, respectively. The units of three coordinate axes are given in Å. These panels display rebound scattering type trajectories which are similar to those of in Figure 5.

**Figure 7**  The long-time interval trajectory calculation up to 500 *fs*. Panel (a) shows the trajectory for the full dimensional trajectory of the H atom (in red) and that of the $B_5$ cluster (in blue). Three representative times *t* = 20, 100, and 280 fs are given with the arrows along the H atom trajectory. Panel (b) is these projection onto the YZ plane. The initial conditions of this calculation were set as ( *R*, α, β , $E_{rel}$ ) = ( 5.5, 72°, 5°, 1.088 ), where *R* is the distance between the H atom and the $B_5$ cluster in the unit of Å, α and β are the azimuthal and the polar angles, respectively. $E_{rel}$ is the initial collision energy in the unit of eV. Panel (c) represents the time-dependence of the relative kinetic energy with the red curve and the internal kinetic energy of $B_5$ with the blue dashed curve. Panel (d) is the full dimensional H atom trajectory with the frozen $B_5$ cluster, and panel (e) is its projection onto the YZ plane. They show a direct rebound scattering. Panel (f) represents the time-dependence of the relative kinetic energy with the red curve and the internal kinetic energy of $B_5$ with the blue dashed curve.

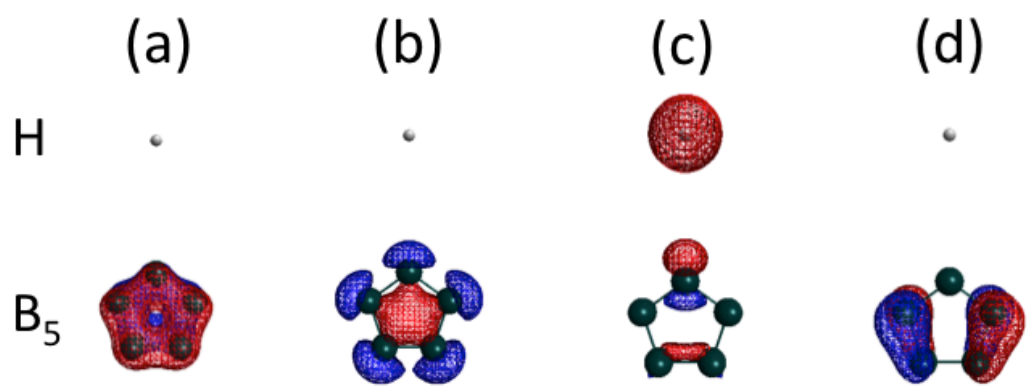

Figure 1

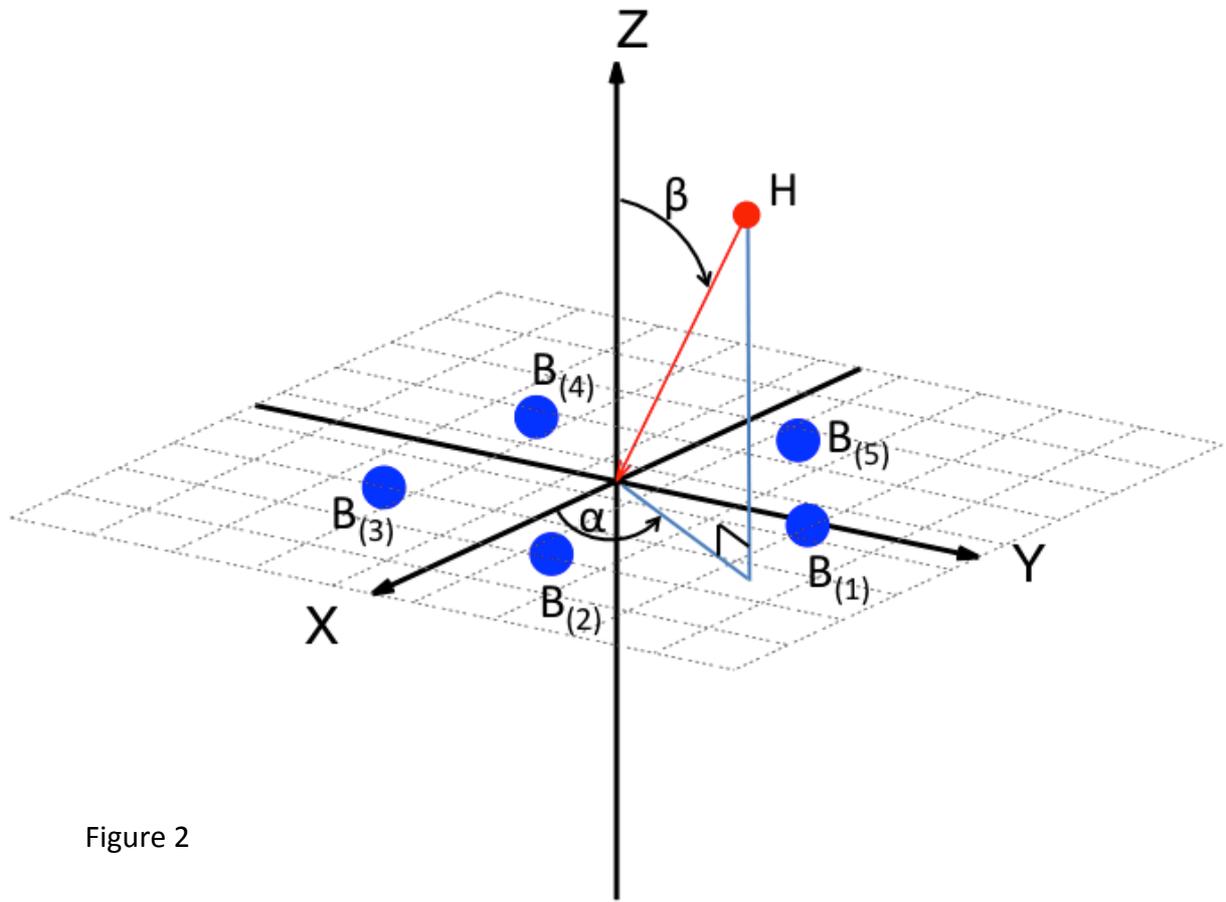

Figure 2

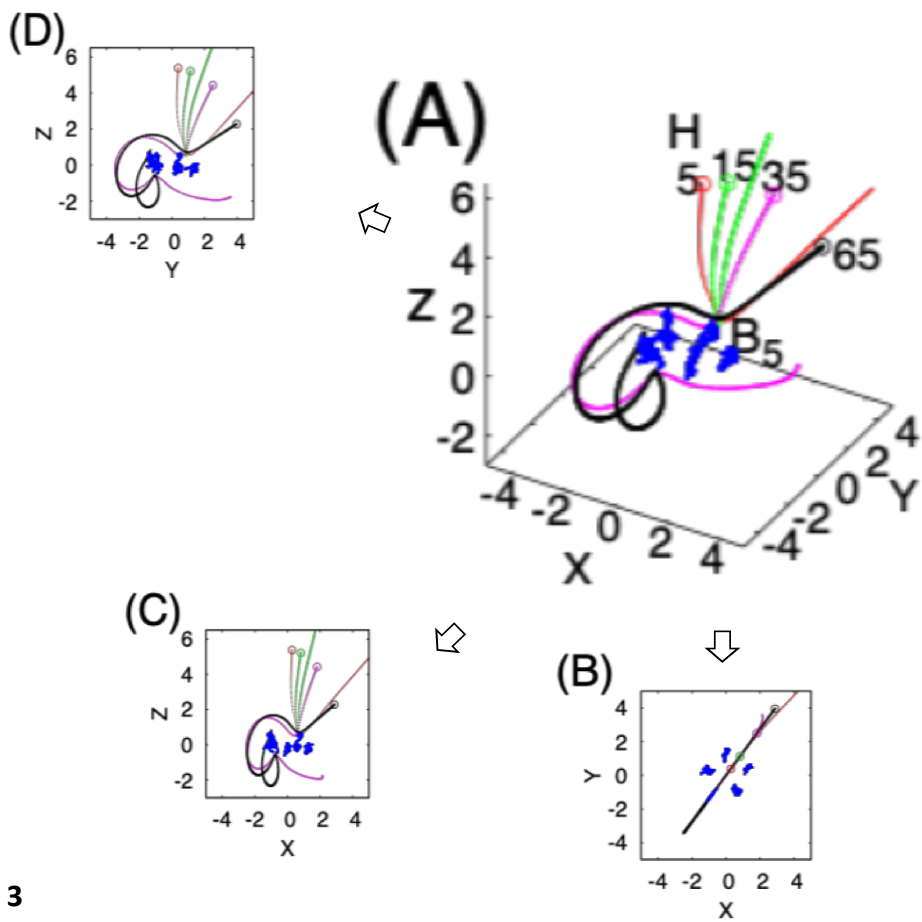

**Figure 3**

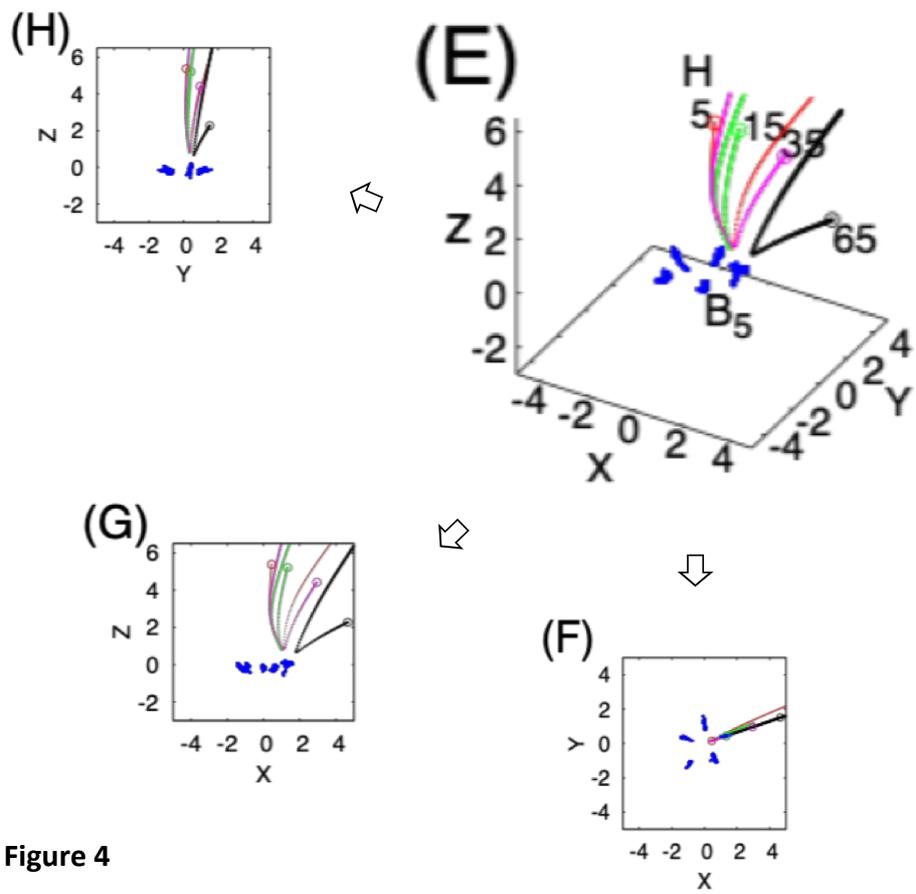

Figure 4

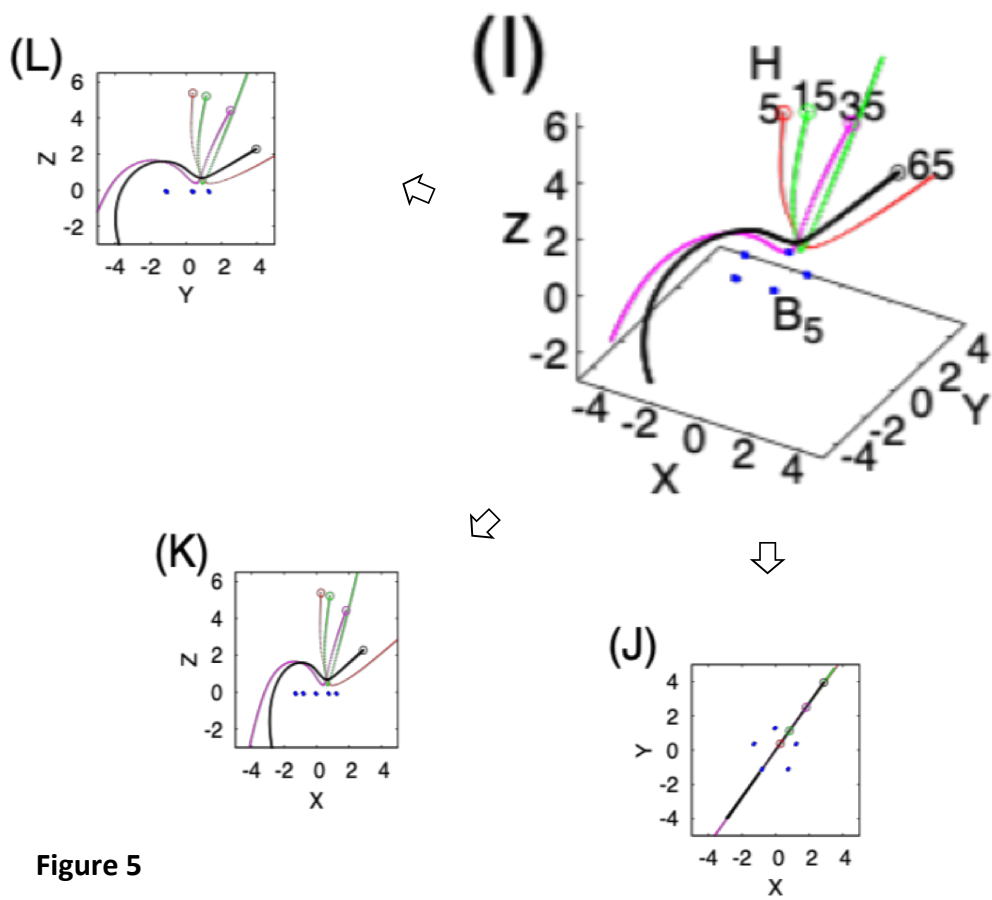

**Figure 5**

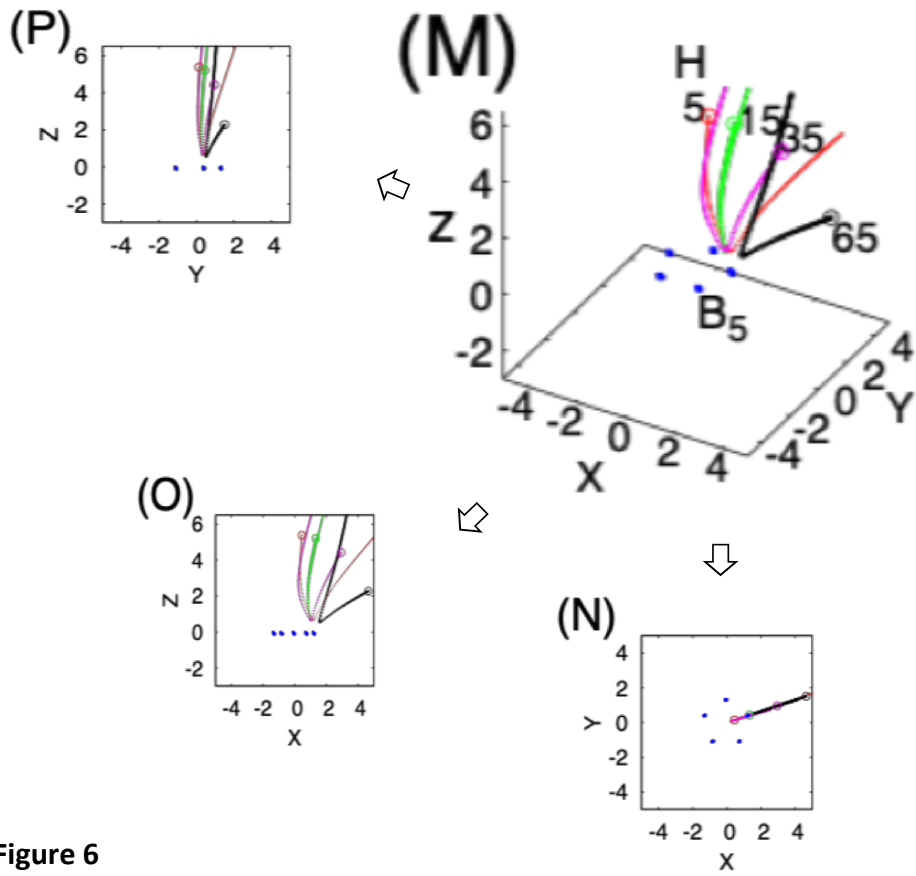

**Figure 6**

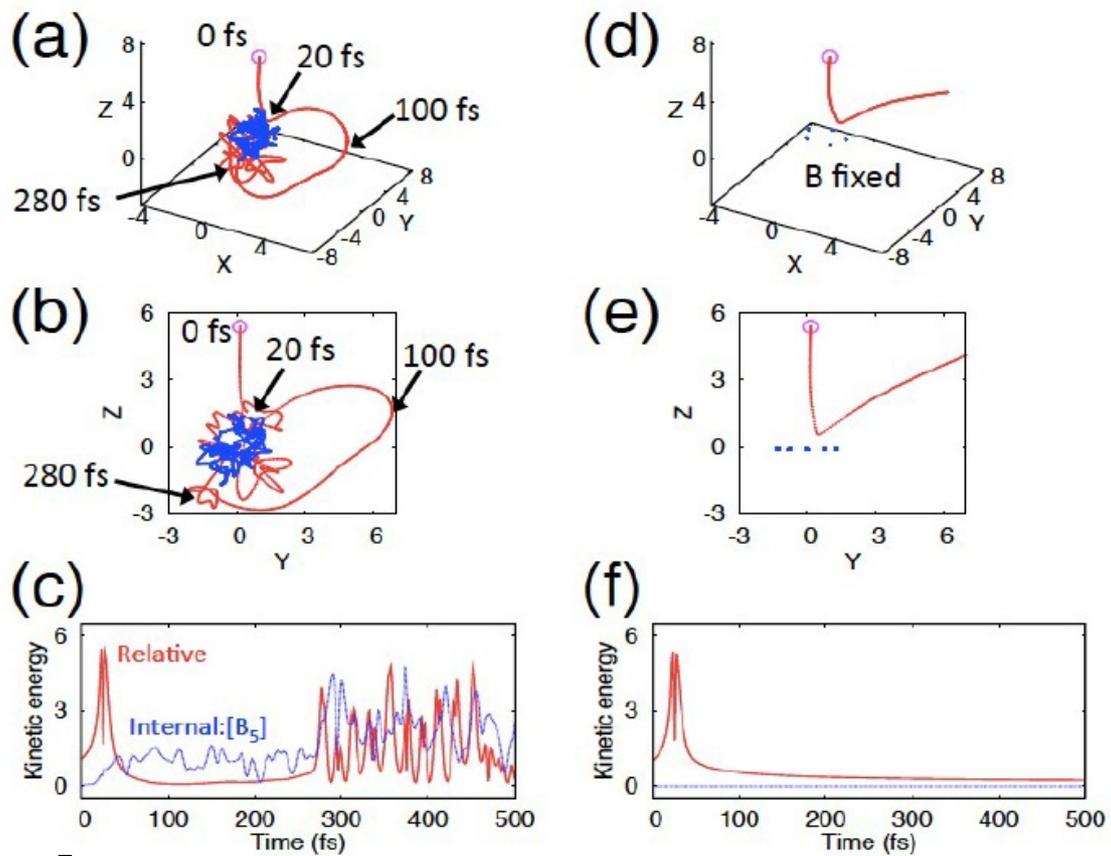

**Figure 7**